\documentclass{ws-procs9x6}

\begin{document}
\title{Correlation Properties of Interfering Electrons
in a Mesoscopic Ring under Nonclassical Microwave Radiation}
\author{D.~I. TSOMOKOS, C.~C. CHONG and A. VOURDAS}
\address{Department of Computing,
School of Informatics,\\
University of Bradford,\\
Bradford BD7 1DP, United Kingdom}
\maketitle

\abstracts{ Interfering electrons in a mesoscopic ring are irradiated with
both classical and nonclassical microwaves. The average intensity of the
charges is calculated as a function of time and it is found that it depends on
the nature of the irradiating electromagnetic field. For various quantum
states of the microwaves, the electron autocorrelation function is calculated
and it shows that the quantum noise of the external field affects the
interference of the charges. Two-mode entangled microwaves are also considered
and the results for electron average intensity and autocorrelation are
compared with those of the corresponding separable state. In both cases, the
results depend on whether the ratio of the two frequencies is rational or
irrational.}

\section{Introduction}
The Aharonov-Bohm effect [1] manifests itself as a nontrivial quantum phase,
whenever electric charges travel in a field-free region enclosing a
magnetostatic flux. This `geometrical phase' has been generalized [2] and the
original results have found applications in various contexts, for example in
conductance oscillations in mesoscopic rings [3] and `which-path' experiments
that use novel solid-state devices [4].

A recent development of these ideas has been to replace the magnetostatic flux
by an electromagnetic field [5]. The objective here is very different, since
this `ac Aharonov-Bohm experiment' constitutes a nonlinear device where the
interaction between the interfering electrons and the photons leads to
interesting nonlinear phenomena [6]. For an overview of related studies on the
interaction of mesoscopic devices with microwaves we refer the reader to [7].

It is interesting to investigate the same phenomena with quantized
electromagnetic fields. This `quantum ac Aharonov-Bohm experiment' with
nonclassical microwaves, has been studied [5] and one can quantify how the
quantum noise destroys slightly the electron interference [8]. The aim is to
investigate how various quantum phenomena and the quantum statistics of the
nonclassical microwaves link to corresponding quantum phenomena on the
electrons.

In what follows we study the interference of the electrons by calculating
their intensity, while they are being irradiated with classical or
nonclassical microwaves. The correlation properties of electron interference
are then studied by calculating the autocorrelation function of the electron
intensity (Sec. 2). We also consider two-mode microwaves with frequencies
$\omega_1$ and $\omega_2$ (Sec. 3). Two-mode microwaves can be factorizable,
separable or entangled [9] and since the problem of entanglement is generally
complex, we approached it using an example. In particular, we assumed that the
two modes of the microwave field form a Bell state and calculated its effect
on electron interference. We found that the result is very different from that
of the corresponding separable case. We conclude in Sec. 4 with a discussion
of our results.

\section{One-mode microwaves}
\subsection{Classical microwaves}
The following system is considered: a beam of electric charges splits into two
possible paths $C_0$ and $C_1$. The charges enter a region that is irradiated
with microwaves (using a suitable waveguide). The microwaves propagate in the
waveguide with the time-dependent magnetic field perpendicular to the plane of
the two paths and the electric field parallel to it. Let $\psi_0$, $\psi_1$ be
the electron wavefunctions with total winding equal to $1$, in the absence of
magnetic field. The effect of the electromagnetic field is the phase factor
$\exp [ie\phi (t)]$ and the intensity is
\begin{equation}
I(t)=|\psi _0+\psi _1\exp [ie\phi (t)]|^2=|\psi _0|^2+|\psi _1|^2 +2|\psi
_0||\psi _1|\Re \{\exp [i(\sigma+e\phi(t))]\}
\end{equation}
where $\sigma = \mbox{arg}(\psi _1)-\mbox{arg}(\psi _0)$. Units in which
$k_B=\hbar=c=1$ are used throughout. For simplicity we consider the case of
equal splitting, in which $|\psi _0|^2=|\psi _1|^2=1/2$ and let $\sigma =0$.
In this case we get
\begin{equation} \label{I_cl}
I(t) = 1 + \cos[e \phi(t)].
\end{equation}

We calculate the autocorrelation function of the electron intensity:
\begin{equation}\label{autocorr}
\Gamma(\tau) = \lim_{T\rightarrow \infty} \frac{1}{2T} \int_{-T}^{T} R(t,\tau)
dt;\;\;\;\;\;\;R(t,\tau)\equiv I(t)I(t+\tau).
\end{equation}
An expansion of $\Gamma(\tau)$ into a Fourier series gives the spectral
density $S_K$:
\begin{eqnarray} \label{psd}
S_K &=& \frac{\Omega}{2\pi}\int_{0}^{2\pi/\Omega}\Gamma(\tau)\exp(-iK\Omega\tau) d\tau \nonumber \\
\Gamma(\tau) &=&\sum_{K=-\infty}^{\infty}S_K\exp(iK\Omega\tau).
\end{eqnarray}

Firstly, we consider the case where the classical time-dependent flux is given
by
\begin{equation}
\phi(t) = \phi_1\sin(\omega t)
\end{equation}
and using Eqs. (\ref{I_cl}) and (\ref{autocorr}) we find the autocorrelation
function:
\begin{equation} \label{G_cl}
\Gamma_{cl}(\tau)=\left[1 + J_{0}(e \phi_1)\right]^2 +
2\sum_{K=1}^{\infty}\left[J_{2K}(e \phi_1) \right]^2 \cos(2K\omega\tau),
\end{equation}
where $J_K$ are Bessel functions. Comparison of Eqs. (\ref{psd}) and
(\ref{G_cl}) shows that $\Omega =2\omega$ and
\begin{equation}
S_0=[1 + J_{0}(e \phi_1)]^2;\;\;\;\;\;\;S_K=[J_{2K}(e \phi_1)]^2.
\end{equation}

\subsection{Nonclassical microwaves}
A monochromatic electromagnetic field of frequency $\omega$ is considered, at
temperatures $k_B T<<\hbar \omega$. We quantize the electromagnetic field by
considering the vector potential $A_i$ and the electric field $E_i$ as dual
quantum variables. The loop $C=C_0-C_1$ is small in comparison to the
wavelength of the microwaves, hence the $A_i$ and the $E_i$ can be integrated
around it and yield the magnetic flux $\phi$ and the electromotive force
$V_{\rm EMF}$, respectively, as dual quantum variables. The annihilation
operator can be introduced as $a=2^{-\frac{1}{2}}\xi^{-1}\left(
\phi+i\omega^{-1}V_{\rm EMF}\right)$, and similarly the creation operator,
where $\xi$ is a constant proportional to the area enclosed by $C$. The flux
operator is consequently written as $\phi(t)= \exp (itH)\phi(0)\exp (-itH)$,
where $H$ is the Hamiltonian that contains the $\omega a^\dagger a$ term and
an interaction term. This interaction term can be neglected for small
currents.

Under these conditions the magnetic flux, which defines the phase factor,
becomes the operator $\hat \phi(t)=(\xi/\sqrt{2}) \left[\exp(i\omega
t)a^\dagger + \exp(-i\omega t)a\right].$ Hence this phase factor
$\exp(ie\phi)$ now is
\begin{equation}
\exp\left[ie\hat \phi(t)\right]=D\left[iq\exp(i\omega t)\right], \ \ \
q=\frac{\xi e}{\sqrt{2}}
\end{equation}
where $D(\lambda)$ is the displacement operator $D(\lambda) = \exp (\lambda
a^\dagger - \lambda ^*a)$. The interference between the two electron beams is
described by the intensity operator
\begin{eqnarray}
\hat{I}(t) = 1 + \cos\left[e \hat{\phi}(t)\right] = 1 + \frac{1}{2}
D\left[iq\exp(i\omega t)\right ] +\frac{1}{2} D\left[-iq\exp(i\omega t)\right
].
\end{eqnarray}
Let $\rho$ be the density matrix describing the external nonclassical
microwaves. The expectation value of the electron intensity is
\begin{equation} \label{I_ave}
\langle I(t) \rangle \equiv \mbox{Tr}\left[\rho\hat{I}(t)\right]=
1+\frac{1}{2}\tilde W(\lambda )+\frac{1}{2}\tilde W(-\lambda ); \ \ \ \
\lambda=iq\exp(i\omega t),
\end{equation}
where $\mbox{Tr}\left[\rho D(\lambda)\right]\equiv \tilde W(\lambda)$ is the
Weyl (or characteristic) function which has been studied by various authors
including ourselves (e.g. [10] and references therein).

\begin{figure}[th]
\centerline{\epsfxsize=3.5in\epsfbox{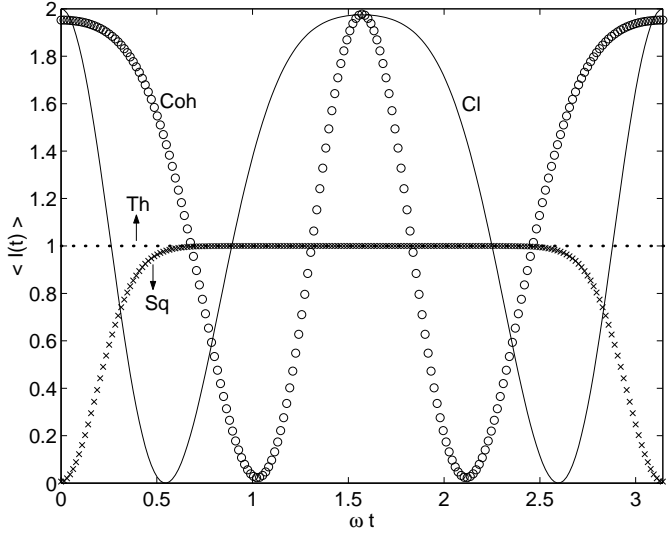}} \caption{$\langle
I(t)\rangle$ as a function of $\omega t$ for $\omega=10^{-4}$, $\langle N
\rangle =200$, $r=6.4$. We use units where $\hbar=k_B=c=1$. Continuous line
represents the case of irradiation with classical microwaves; line of circles,
coherent states; line of crosses, squeezed states; and dotted line, thermal
states.}
\end{figure}

We have calculated $\langle I(t) \rangle$ for various quantum states of the
microwaves (using results for $\tilde W(\lambda)$ in Ref. [11]). In order to
find the $\Gamma(\tau)$ from Eq. (\ref{autocorr}), one needs to calculate the
quantity
\begin{equation} \label{R_exp}
R(t,\tau)\equiv \mbox{Tr}\left[\rho\hat{I}^{\dagger}(t)\hat{I}(t+\tau)\right]
\end{equation}

\begin{figure}[th]
\centerline{\epsfxsize=3.5in\epsfbox{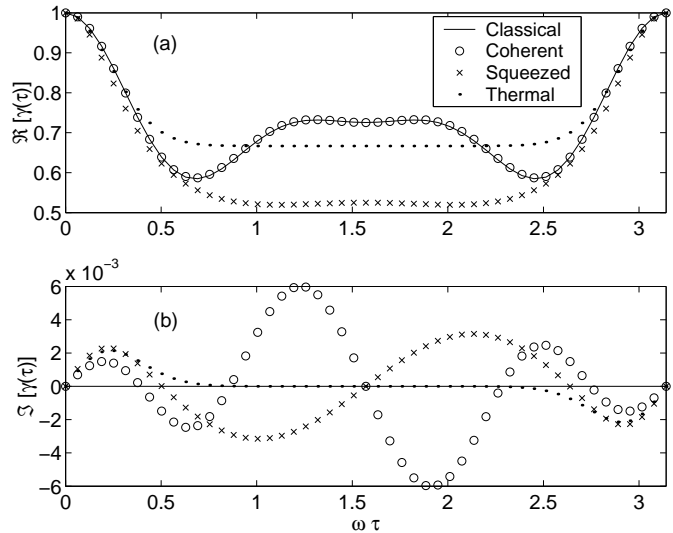}}
\caption{$\gamma(\tau)$ as a function of $\omega \tau$ for $\omega=10^{-4}$,
$\langle N \rangle =200$, $r=6.4$. Part (a) shows the real part of
$\gamma(\tau)$; part (b) shows the imaginary part. We use units where
$\hbar=k_B=c=1$.}
\end{figure}

Numerical results are presented for different quantum states that we
calculated. In particular, we plot four cases: classical microwaves and
nonclassical microwaves in coherent, squeezed, and thermal states. For a
meaningful comparison, we consider the case where the average number of
photons $\langle N \rangle$ in coherent, squeezed, and thermal states is the
same:
\begin{eqnarray}
\langle N \rangle =|A|^2&=& \left[\sinh \left(\frac{r}{2}\right)\right]^2 +
\left[\cosh \left(\frac{r}{2}\right)
- \sinh\left(\frac{r}{2}\right) \right]^2 B^2 \nonumber\\
&=&\frac{1}{\exp(\beta \omega) -1}.
\end{eqnarray}
For the classical case we took $\phi_1^2 = 2|A|^2=2\langle N\rangle$. In all
results of Figs. 1 to 3, $\omega=10^{-4}$ (which in our units is $eV$),
$\langle N \rangle =200$, $r=6.4$.

\begin{figure}[th]
\centerline{\epsfxsize=3.5in\epsfbox{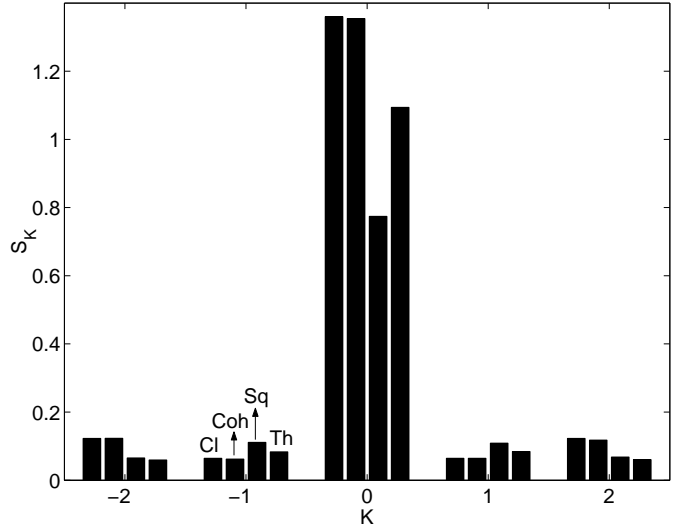}} \caption{$S_K$
coefficients for the electrons with $\langle N \rangle =200$, $r=6.4$. We use
units where $\hbar=k_B=c=1$. The bars correspond to irradiation of the ring by
(from left to right): classical, coherent, squeezed, and thermal states.}
\end{figure}

The results show that the quantum noise in the irradiating microwaves affects
the electron interference. All microwaves that we have considered have the
{\bf same average number of photons}, but differ in the quantum noise. These
four types of microwaves lead to different electron interference results and
different autocorrelation functions. Irradiation of the electrons by
nonclassical microwaves leads to nonzero value of the imaginary part of the
electron autocorrelation function. This is not so (i.e. the imaginary part of
$\Gamma(\tau)$ vanishes) when the ring is irradiated with classical
microwaves.

\section{Two-mode nonclassical microwaves}
We consider two-mode nonclassical microwaves. We are particularly interested
to study how entangled two-mode microwaves affect the electron interference.
For this reason we consider a Bell state $|s\rangle =2^{-1/2}(|01\rangle
+|10\rangle )$ where $|01\rangle$ , $|10\rangle$ are two mode number
eigenstates. For comparison we also consider the separable (disentangled)
state
\begin{equation} \label{sep}
\rho _{sep}=\frac {1}{2} (|01\rangle \langle 01| +|10\rangle \langle 10|).
\end{equation}
Clearly, the density matrix of the entangled state $\rho _{ent}=|s\rangle
\langle s|$ can be written as
\begin{equation} \label{ent}
\rho _{ent}=\rho _{sep}+\frac {1}{2} (|01\rangle \langle 10| +|10\rangle
\langle 01|).
\end{equation}
In this case the phase factor $\exp[ie\phi(t)]$ becomes the product of two
displacement operators and, consequently, the intensity becomes
\begin{eqnarray}
\hat{I}(t) = 1 + \frac{1}{2} D_1(\lambda_1)D_2(\lambda_2)+ \frac{1}{2}
D_1(-\lambda_1)D_2(-\lambda_2); \ \lambda_j = iq\exp(i\omega_j t)
\end{eqnarray}
for two modes ($j=1,2$). Therefore, we find that
\begin{eqnarray}
\langle I(t)\rangle_{sep}&=& 1 + \left(1-q^2\right)\exp\left(-q^2\right), \\
\langle I(t)\rangle_{ent}&=& \langle I(t)\rangle_{sep} -
q^2\exp\left(-q^2\right)\cos[(\omega_1 -\omega_2)t].
\end{eqnarray}

It is seen that for this example, the $\langle I(t)\rangle _{sep}$ is constant
in time, while the $\langle I(t)\rangle_{ent}$ is an oscillatory function of
time. Clearly, different correlations among the two irradiating modes of the
microwaves may lead to different average electron intensities.

\section{Discussion}
The subject of mesoscopic devices interacting with microwaves has received
attention in the last few years (e.g., Ref. [7]). Our contribution has been to
consider that these microwaves are prepared in various nonclassical states
[5,8,12]. Here we have quantified the effect of the quantum noise on electron
interference. More specifically we have calculated both the electron average
intensity and the spectral density for several types of nonclassical
microwaves and a comparison of the results with the case of classical
microwaves (Figs. 1-3), demonstrates clearly that the presence of both
classical and quantum noise in the nonclassical microwaves affects the
electron intensity. What is more, when the ring is irradiated with two-mode
microwaves, then entanglement among these two modes (i.e., the formation of a
Bell state) leads to a time-dependent expectation value of the electron
intensity.



\begin{thebibliography}{0}
\bibitem{1} M. Peshkin and A. Tonomura, \emph{The Aharonov-Bohm effect}, Lecture notes in Physics Vol. 340,
             Berlin: Springer (1989).

\bibitem{2} A. Shapere and F. Wilczek (ed), \emph{Geometric Phases in Physics},
            Singapore: World Scientific (1989).

\bibitem{3} S. Washburn and R.A. Webb, {\it Adv. Phys.} {\bf 35}, 375 (1986)\\
            A.G. Aronov and Y.V. Sharvin, {\it Rev. Mod. Phys.} {\bf 59}, 755 (1987).

\bibitem{4} G. Hackenbroich, {\it Phys. Rep.} {\bf 343}, 464 (2001).

\bibitem{5} A. Vourdas, {\it Phys. Rev.} {\bf B54}, 13175 (1996) \\
            A. Vourdas and B.C. Sanders, {\it Europhys. Lett.} {\bf 43}, 659 (1998).

\bibitem{6} M.P. Silverman, {\it Nuovo Cimento} {\bf B97}, 200 (1987)\\
            M. Buttiker, {\it Phys. Rev.} {\bf B46}, 12485 (1992).

\bibitem{7} M. Buttiker, {\it J. Low Temp. Phys.} {\bf 118}, 519 (2000)\\
            R. Deblock \emph{et. al.}, {\it Phys. Rev.} {\bf B65}, 075301 (2002).

\bibitem{8} P. Cedraschi, V.V. Ponomarenko, M. Buttiker, {\it Phys. Rev. Lett.} {\bf 84}, 346 (2000)\\
            A. Vourdas, {\it Phys. Rev.} {\bf A64}, 053814 (2001).

\bibitem{9} R.F. Werner, {\it Phys. Rev.} {\bf A40}, 4277 (1989);
        A. Peres, {\it Phys. Rev. Lett.} {\bf 77}, 1413 (1996);
        R. Horodecki and M. Horodecki, {\it Phys. Rev.} {\bf A54}, 1838
        (1996);
        V. Vedral \emph{et. al.}, {\it Phys. Rev. Lett.} {\bf 78}, 2275 (1997).

\bibitem{10} S. Chountasis and A. Vourdas, {\it Phys. Rev.} {\bf A58}, 848 (1998).

\bibitem{11} A. Vourdas, {\it Phys. Rev.} {\bf B49}, 12040 (1994).

\bibitem{12} C.C. Chong, D.I. Tsomokos, A. Vourdas, {\it Phys. Rev.} {\bf A66}, 33813 (2002).

\end{thebibliography}
\end{document}